# EagleTree: Exploring the Design Space of SSD-Based Algorithms

Niv Dayan[1], Martin Kjær Svendsen[1], Matias Bjørling[1], Philippe Bonnet[1], Luc Bouganim[2,3]

[1] IT University of Copenhagen
Copenhagen, Denmark
{mabj,phbo,nday}@itu.dk

[2] INRIA Paris-Rocquencourt
Le Chesnay, France
Luc.Bouganim@inria.fr

[3] PRISM Laboratory
Univ. of Versailles, France
Luc.Bouganim@prism.uvsq.fr

## ABSTRACT

Solid State Drives (SSDs) are a moving target for system designers: they are black boxes, their internals are undocumented, and their performance characteristics vary across models. There is no appropriate analytical model and experimenting with commercial SSDs is cumbersome, as it requires a careful experimental methodology to ensure repeatability. Worse, performance results obtained on a given SSD cannot be generalized. Overall, it is impossible to explore how a given algorithm, say a hash join or LSM-tree insertions, leverages the intrinsic parallelism of a modern SSD, or how a slight change in the internals of an SSD would impact its overall performance. In this paper, we propose a new SSD simulation framework, named EagleTree, which addresses these problems, and enables a principled study of SSD-Based algorithms. The demonstration scenario illustrates the design space for algorithms based on an SSD-based IO stack, and shows how researchers and practitioners can use EagleTree to perform tractable explorations of this complex design space.

## 1. INTRODUCTION

Flash-based Solid State Drives (SSDs) offer the same block device interface as hard disk drives (HDDs). It is thus seemingly transparent for a database administrator to replace HDDs by SSDs. The problem is that SSDs do not respect the performance contract that has always been valid for HDDs: e.g., sequential IOs are no longer orders of magnitude faster than random IOs. Worse, there is no consistent performance contract that all SSDs adhere to. We have argued in [2] that this absence of a performance contract is due to the high level of software and hardware complexity, throughout the portion of the IO stack hidden behind the block device interface.

So, even if SSDs look like HDDs, they behave in a very different way. In fact, the behavior of a given SSD is hard to characterize as it depends on undocumented internal features at the hardware level, e.g., SSD geometry, underlying flash chips, and mainly at the software level, e.g., the Flash Translation Layer (FTL) embedded on the SSD controller.

The fact that complex SSDs internals are hidden and undocumented raises problems, not just in terms of experimental methodology – as care must be taken to bring an SSD to a well defined state before running experiments in order to obtain reproducible results [3] – but essentially in terms of system design, as it is impossible for practitioners and researchers to explore how SSD internals impact overall system performance. More specifically, we seek to explore the following questions that have received little attention so far:

- How does SSD parallelism impact performance (or the dual question: how can an algorithm efficiently leverage SSD parallelism)? A flash-based SSD contains tens to hundreds of flash chips wired in parallel to the SSD controller through multiple channels. The SSD scheduler deals with parallelism across and within flash chips and must decide *which* IO should be scheduled, and for writes, *where* (i.e., on which LUN[1]) it should be done, and precisely *when* (with respect to other IOs).

- How do garbage collection and wear leveling interfere with application IOs? Each update leaves an obsolete flash page (with a before image). Over time, obsolete flash pages accumulate, and are reclaimed through garbage collection (GC). In addition, the FTL relies on wear leveling (WL) to distribute the erase count across flash blocks and mask bad blocks. Note that both GC and WL read live pages from a victim block and write those pages at other locations, before that block is erased. As a result, GC and WL interfere with the application's IOs, possibly compromising throughput and contributing to latency variability. Note that application IOs also interfere with each other, which raises issues of fairness that have not been properly addressed so far.

- What is the impact of replacing the block layer by a communication interface, hopefully better suited to deal with the complexity of SSDs? We have argued for such a cross layer approach [2], but we cannot rely on existing SSDs to experiment with its design.

In order to study these questions, we must resort to simulations where SSD internals can be manipulated with controlled experiments. Existing SSD simulators fall short of this requirement. Early efforts such as FlashSim [7] (open source) and SSDSim [1] (Microsoft license) are inadequate at the hardware level: they fail to represent SSD parallelism and chip characteristics. While the more recent open source simulator NANDFlashSim [6] improves the hardware and parallelism representation, it does not expose a design space at the controller level that allows for rapid experimentation with mapping, GC, WL or scheduling policies.

---

[1] Flash chips are decomposed into logical units, denoted LUNs that constitute the minimum granularity of parallelism. This notion of LUN, introduced in the Open Nand Flash Interface (ONFI) standard, abstracts away the notions of packages, chips and dies.

In this paper, we present EagleTree, an open source software simulation framework for SSD-based applications. EagleTree encompasses the complete IO stack: application, operating system, SSD controller and flash chips array. EagleTree exposes a large, complex design space (see Section 2), which can be explored in a tractable way. The demonstration scenario (see Section 3) illustrates the complexity of this design space and shows how researchers and practitioners can use EagleTree to explore interplays between parameters across the IO stack.

## 2. THE EAGLETREE SIMULATOR

### 2.1 Overview

EagleTree is an open source simulator (available at https://github.com/ClydeProjects/EagleTree). It does not only simulate an SSD but also (a subset of) the OS and applications utilizing it. The first advantage of this approach is that EagleTree is an entire system operating in virtual time. It is thus possible to conduct large and complex design-space explorations, involving hundreds of experiments, in a tractable way. A second advantage is that we can experiment with cross-layer designs.

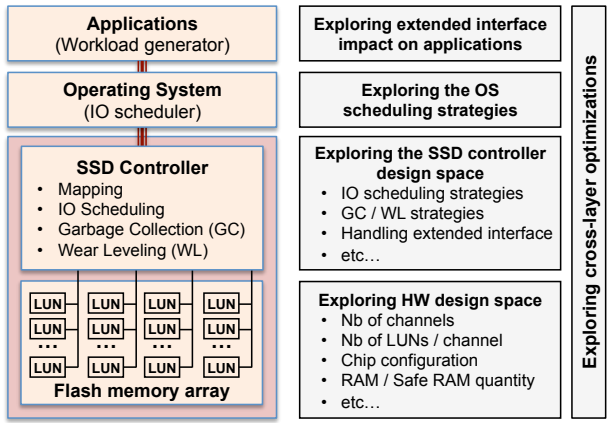

**Figure 1: EagleTree architecture and design space exploration**

EagleTree is composed of four layers, from the bottom-up: the hardware, the SSD controller, the operating system, and the application layers (see Figure 1). With EagleTree we can study questions related to an individual layer, or we can conduct cross-layer studies. For example:

*At the hardware layer:* Where are the hardware bottlenecks? What is the impact of the underlying flash chips characteristics? How should we use advanced commands (e.g. copybacks, pipelining), and what trade-offs is their usage subject to?

*At the SSD controller layer:* What is the impact of the mapping strategy? When should we trigger garbage collection or wear leveling? How should we schedule application reads and writes (1) relative to each other and (2) relative to internal IOs? What is the impact of different interleaving strategies? What is the best usage for RAM or for battery-backed RAM? What is the impact of applications' IOs queue size?

*At the OS scheduler level:* What is the best scheduling strategy (e.g., FIFO, CFQ, priorities)? How many outstanding IOs should be submitted to the SSD? What useful meta-information may be transmitted to the SSD in case of cross layer optimizations?

*At the application layer:* How can an algorithm leverage SSD internal parallelism? How should we submit synchronous and asynchronous IOs?

Finally, considering a *cross-layer approach*: How can we leverage an open interface to cooperate with the OS and the SSD? What is the impact on application algorithms (e.g., indexing, extent-based space allocation, join algorithms, external sorting algorithms)? How should work be divided across layers (e.g., scheduling at OS level or SSD level or both)?

The obvious drawback of our approach is that the entire system is simulated, and thus, far away from a deployable system. Our goal is that EagleTree should be useful to answer some basic questions, but we recognize that the answers that we obtain will have to be confronted to actual system components. We are exploring an approach where the fully simulated environment of EagleTree is complemented by (1) a software-based SSD simulator (based on the hardware and controller layers of EagleTree) connected to a real OS and real applications -- the problem is then to store the manipulated data in RAM (otherwise it may slowdown the execution), and to map the virtual IO time to real time, and (2) a hardware-based SSD simulator using the OpenSSD hardware [10]. We think that these approaches, more realistic but less agile [9], are best suited in a second phase, once EagleTree has been used to get a basic understanding of the key trade-offs.

### 2.2 Design and Implementation

Each EagleTree layer consists of configurable parameters and customizable policies. From the bottom-up:

**Hardware:** EagleTree allows users to set up every hardware parameter of the simulated SSD: basic flash chip timings (i.e., to send a command, transfer data on a channel, read, write or erase), flash chip and SSD geometry. Moreover, EagleTree allows specifying the flash chip type (i.e., SLC or MLC) and its support for advanced commands. Finally, EagleTree includes a memory manager used to track the amount of RAM and battery-backed RAM used for the controller's metadata and IO buffers. All these parameters are variables that can be set, viewed and updated with ease. Predefined configurations are provided based on existing SSDs and flash chip datasheets.

**SSD Controller:** The SSD controller is responsible for orchestrating mapping, garbage-collection, wear leveling modules and scheduling.

The **mapping** scheme supports the virtualization of the address space, mapping logical addresses onto physical ones. For reads, the mapping scheme must look up the physical address corresponding to the logical address of the incoming IO. For writes, the mapping scheme imposes constraints on which physical address a given IO might be bound to. As a result, the mapping scheme potentially restricts the scheduling policy. For now, we have considered the most flexible schemes i.e., page-based mappings: the well-known DFTL [5] and a page-based mapping scheme where the entire mapping is kept in RAM.

For page-mapping FTLs, **garbage-collection** should fulfill the following requirements. First, it is desirable to wait as long as possible before performing garbage-collection. Doing so maximizes the number of invalid pages across the SSD, thereby ensuring that victim blocks have few live pages. On the other hand, GC must not occur so late that the FTL actually runs out of available space for incoming writes. Second, it is desirable to maintain free space on every LUN to maximize the flexibility for where writes can be made. The default GC module strives to fulfill these goals by triggering GC so that a given number of blocks (*GC Greediness* parameter) are always free on each LUN.

The default **wear leveling** module keeps track of (1) the ages of all blocks, (2) a timestamp for each block marking the time in which it was last erased, (3) the average length of time it takes a block to be erased, and (4) the current time. Using this information, the WL module can identify particularly young blocks that have not been erased for a very long time, and can target them for static wear leveling. Moreover, EagleTree contains dynamic wear-leveling strategies that allow maintaining several free blocks of different ages in each LUN. The overall goal is to associate hot data with young blocks and cold data with old blocks. Temperature detection for pages can be done by (1) assuming the pages migrated in static wear-leveling are cold, and everything else is hot, or (2) using a temperature detection mechanism for each page such as the one described in [8], which we have implemented, or (3) using information about the temperature of data coming through an open interface from the application.

EagleTree supports parallelism among channels and operation interleaving within a channel thus making the IO scheduler a central part in EagleTree. With regards to **IO scheduling**, the problem is roughly the following. Given the state of the flash chip array and a queue of pending IOs from various sources (e.g. application, garbage-collection, mapping, etc.), of various types (e.g. read, write, erase, copy-back), and that have been waiting in the queue for different lengths of time, which IO should be executed next and where? EagleTree provides a modular framework for exploring scheduling policies. For example, it is possible to:

- Control when and where to trigger internal operations, such as garbage-collection, wear-leveling;
- Implement priority schemes that differentiate among IOs based on their sources, types, and waiting times;
- Set deadlines for different IO types, and control the way in which overdue IOs are handled as relative to other IOs;
- Record and exploit information about logical address patterns (e.g. sequential vs random, hot vs cold);
- Control the aggressiveness of interleaving and copy-back operations;

Obviously, these defaults strategies can be overwritten or complemented. In addition, other modules can be added to the SSD controller, e.g., a write-buffering module that uses battery-backed RAM to temporarily store data before it is written on flash pages.

**Open Interface:** EagleTree takes a departure from the traditional block device interface by basing communication between the OS and the SSD on an extensible messaging framework that allows the operating system and SSD to communicate as peers. Users are able to create new types of messages between the SSD and the OS conveying any amount of information or instructions. Using this framework, it is possible to build arbitrarily complex communication protocols between the SSD and OS. Investigating what we stand to gain from more explicit communication between the OS and SSD is the subject of our ongoing work. For now, we hope to inspire and excite the reader by sketching a few examples and their potential impact on performance.

- *Priorities:* the OS can communicate to the SSD the priority of an IO. The SSD can take this into account by offering the IO special treatment in terms of scheduling.
- *Update-locality:* the OS can inform the SSD which pages share update-locality. The SSD can then write these pages so as to minimize subsequent garbage-collection.
- *Temperatures:* the OS can inform the SSD whether the page being written is likely to be updated soon. The SSD can use this to benefit wear-leveling and garbage-collection efficiency.

**OS Scheduler:** The Operating System manages IO requests incoming from multiple simulated concurrent threads. It maintains a pool of pending IOs from each thread and decides, based on a customizable scheduling policy, which IOs to issue next to the SSD. This policy can take into account the IO type (e.g. read/write/trim), its priority, the dispatching thread, etc. The default scheduling strategy is FIFO. Once the SSD has completed executing an IO, it interrupts and notifies the OS. The OS then activates the thread that dispatched the IO. The thread can respond by issuing additional IOs.

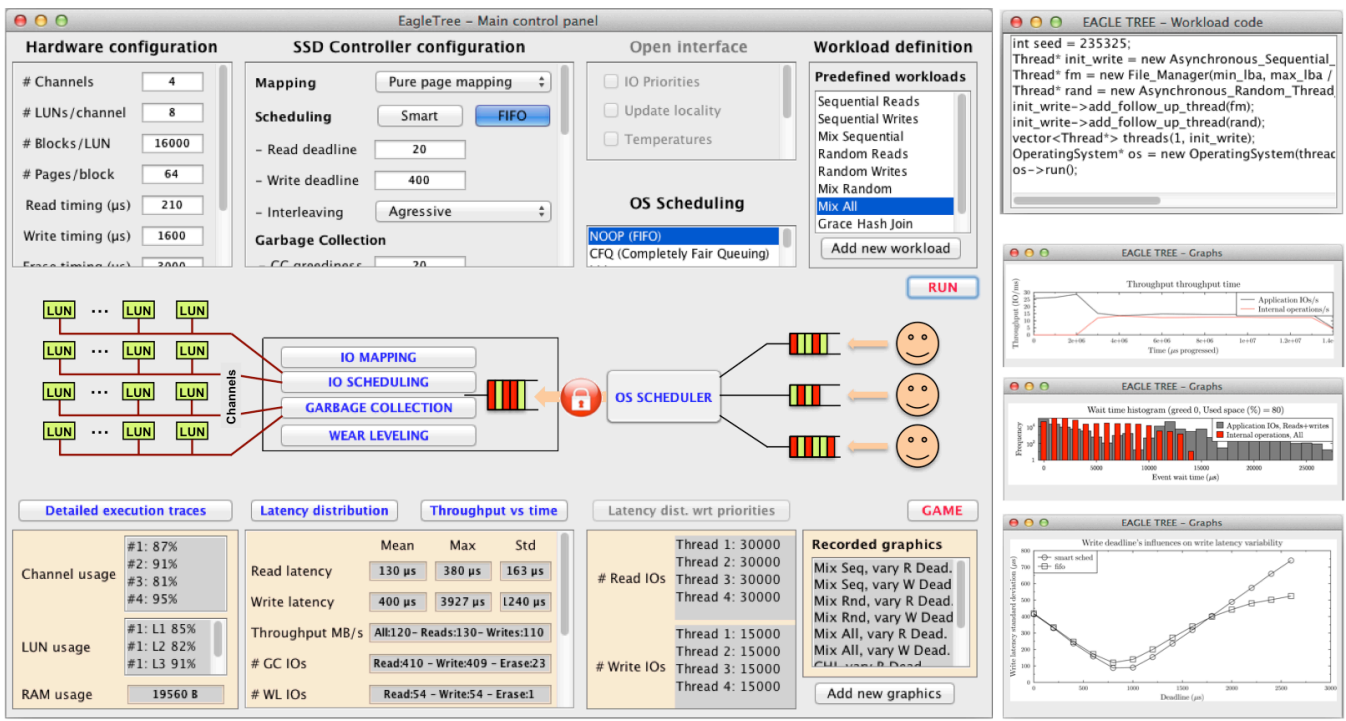

Figure 2: EagleTree demonstration main window (on the left) and example of code and results

**Threads:** The Thread layer is a programming framework that gives users absolute control over the workload. Users are able to extend an abstract thread class by providing a definition for two methods: init() and call_back(). The init() method is called by the OS when the thread is initialized, and the call_back() method is triggered every time an IO originating from that thread completes. Within each of these methods, it is possible to issue any number of messages to the OS. A user can create any number of threads with arbitrarily complex behaviors. For example, we have implemented threads simulating the behavior of a file system as well as a thread that follows the IO pattern of Grace hash Join.

## 2.3 Running Experiments

EagleTree contains an experimental suite API, which consists of experiment templates. An experiment template takes (1) an SSD parameter or policy (2) a strategy for how to vary it in an experiment, and (3) a workload definition. It runs an experiment and produces a comprehensive amount of visual statistical output. This includes graphs showing how performance metrics (e.g., throughput, latency, latency variability) evolved with respect to the given parameter or policy, as well as graphs showing how various metrics evolved across time in the experiments, and massive visual traces showing exactly how every IO was handled throughout the simulator components. EagleTree contains other useful features to allow conducting controlled, repeatable experiments. For example, it is possible to attach statistics gathering objects to an individual thread to measure its performance. It is also possible to create dependencies among threads. This latter feature is particularly useful for bringing the SSD to a well-defined state. This can typically be done by starting thread(s) that write over the entire logical address space sequentially and/or randomly [4] and then triggering the experiment workload once the preparation threads finished, and measuring performance only for the experiment workload.

## 3. DEMONSTRATION SCENARIO

The purpose of the demonstration is to show through experiments, that: (1) the design space of SSD-based algorithms is vast; that (2) opening the interface is interesting but still increases that design space; that (3) EagleTree can help to explore that design space; and finally that (4) interesting solutions are sometime counter-intuitive. The demonstration will be centered on scheduling issues and will roughly follow the following outline.

**Layered Tour:** This part uses the demonstration GUI shown on Figure 2. We will first introduce the approach and EagleTree architecture. Then, attendees will choose configuration parameters (e.g. hardware setup, controller and OS policies, application workload) then run the simulator, and observe live results in terms of numerical performance metrics, traces, and graphical outputs (lower part of the GUI and right part of Figure 2). We will also show pre-computed graphs using longer experiments to show the full power of EagleTree's experimental suite. We will pay particular attention to the impact of scheduling policies on performance, and explain why prioritizing between application reads and writes is not always easy. We will introduce the challenge of scheduling internal operations as non-obtrusively as possible.

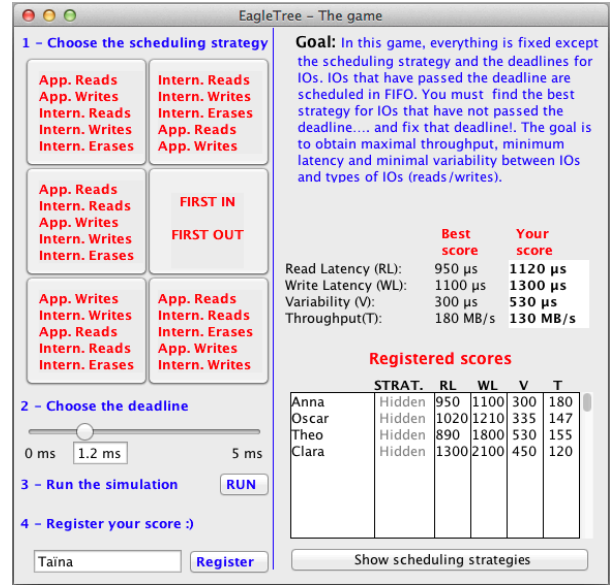

**Figure 3: EagleTree game window**

**Open Interface Appetizers:** In a second phase, we will open the block device interface. By unlocking the red lock shown on the figure, we will enable the attendee to choose from a range of possible extensions to the block device interface (e.g., explicit communication of locality, data temperature), guiding the attendee such that large variation in performance can be observed.

**Game:** We will close the demonstration proposing a game to the attendee. Using the second GUI (see Figure 3) and knowledge acquired during the presentation, the user will have to guess the optimal combination of scheduling policies given a subset of the SSD scheduling design space. The attendee's objective will be to maximize throughput for a given workload while balancing mean latency and latency variability between different types of IOs. The user who draws nearest to the optimal configuration at each demo session will win an EagleTree Tshirt.